\documentclass[twocolumn,showpacs,preprintnumbers,amsmath,amssymb]{revtex4}
\usepackage{graphicx}
\usepackage{dcolumn}
\usepackage{bm}

\begin{document}

\preprint{APS/123-QED}

\title{A measure of the impact of future dark energy experiments based
  on discriminating power among quintessence models. }

\author{Michael Barnard}
\author{Augusta Abrahamse}
\author{Andreas Albrecht}
\author{Brandon Bozek}
\author{Mark Yashar}
 \affiliation{Department of Physics, One Shields Ave.; University of California; Davis, CA 95616}

\begin{abstract}
We evaluate the ability of future data sets to discriminate among {\em
  different} quintessence dark
energy models.  This approach gives an alternative (and complementary)
measure for assessing the impact of future experiments, as compared
with the large body of literature that compares experiments in abstract
parameter spaces (such as the well-known $w_0-w_a$ parameters) and more
recent work that evaluates the constraining power of experiments on individual
parameter spaces of specific quintessence models.   We use the Dark
Energy Task Force (DETF) models of future data sets and compare the
discriminative power of experiments designated by the DETF as Stages
2, 3, and 4 (denoting increasing capabilities).  Our work reveals
a minimal increase in discriminating power when comparing Stage 3 to
Stage 2, but a very striking increase in discriminating power when
going to Stage 4 (including the possibility of completely eliminating
  some quintessence models).  We also see evidence that even modest improvements
over DETF Stage 4 (which many believe are realistic) could result in
even more dramatic discriminating power among quintessence dark energy
models. We develop and demonstrate the technique of using the
independently measured modes of the equation of state (derived from
principle component analysis) as a common parameter space in which to
compare the different quintessence models, and we argue that this
technique is a powerful one.
We use the PNGB,
Exponential, Albrecht-Skordis, and Inverse Tracker (or Inverse Power
Law) quintessence models for this work.  One of our main results is
that the goal of discriminating among these models sets a concrete
measure on the capabilities of future dark energy experiments.
Experiments have to be somewhat better than DETF Stage 4 simulated
experiments to fully meet this goal.
\end{abstract}


\maketitle

\section{Introduction}
\label{sec1}

Over the last decade or so there has been mounting evidence that the
energy of the universe has a large accelerating component,
dubbed ``dark energy''\cite{Frieman:2008sn}.  As the evidence has become more convincing,
there has been growing enthusiasm for launching a major program to
collect additional data that will help us better understand the nature
of the dark energy \cite{Albrecht:2007xq}, and indeed considerable
progress is being made on this front.

To plan a strong program of dark energy studies one needs to assess
the relative impact of different possible experiments.  This has most
often been done by describing the dark energy in some abstract
parameter space and calculating how much a given data set could
constrain those abstract parameters.  For example, the Dark Energy Task
Force (DETF) \cite{Albrecht:2006um}, building on earlier work
\cite{Linder:2002et,Huterer:2000mj}, used a standard two parameter model of the
dark energy equation of state $w$ as a function of
cosmic scale factor $a$ given by $w(a) = w_0+w_a(1-a)$ to form a figure
of merit based on constraining power in the $w_0-w_a$ parameter
space.  For the most part, other authors have used other abstract dark
energy parameterizations
\cite{Albrecht:2006um,Huterer:2006mv,Albrecht:2007qy}, but more recently we have
extensively explored the impact of future experiments using the
constraints produced on the actual parameters of scalar field dark
energy models
\cite{Barnard:2007ta,Abrahamse:2007te,Bozek:2007ti,Yashar:2008xx}.
That work gives another window on the power of future
experiments, which we have argued is largely consistent with the DETF results in the
$w_0-w_a$ parameter space.

Our recent work \cite{Barnard:2007ta,Abrahamse:2007te,Bozek:2007ti,Yashar:2008xx}
shows the constraining power of
future experiments on specific dark energy models.  However, because
the natural parameter space of each quintessence model is very
different from the others, we were not able to use our techniques
to directly evaluate the ability of experiments to favor one
dark energy model strongly over another. The one exception to this is
``cosmological constant'' dark energy (which has $w(a) = -1$).  Each
quintessence model we considered had a part of parameter space where
the quintessence closely mimicked a cosmological constant, and we used that
fact systematically in \cite{Barnard:2007ta,Abrahamse:2007te,Bozek:2007ti,Yashar:2008xx} to consider discriminating
power in the quintessence vs. cosmological constant domain.

This paper builds on that earlier work to make a more comprehensive
analysis of the ability of future data to discriminate among different
quintessence models.  The key new ingredient is the use of a specially
chosen parameter space to represent the different quintessence models
in a common and comparable form.  To this end we use the ``independently
measured modes'' of $w(a)$, which have long been appreciated for a
variety of reasons
\cite{Huterer:2000mj,Huterer:2002hy,Knox:2004vw,Linder:2005nh,Riess:2006fw,Albrecht:2007qy,dePutter:2007kf,Sullivan:2007tx}.
The modes, which
are different for each experiment, represent the modes or ``moments'' of $w(a)$
of which uncorrelated measurements are made by that particular experiment.  This
feature allows us to identify the modes which are best measured and
analyze them in a straightforward way (due to the lack of
correlations).  These modes comprise a basis which spans the space of
possible functions $w(a)$. A given quintessence model with specific
fixed parameters will give a specific function $w(a)$ which can then
be expanded in the modes, and the expansion coefficients form the
parameter space in which we work.

We use the PNGB \cite{Frieman:1995pm,Abrahamse:2007te},
Exponential \cite{Ratra:1987rm,Bozek:2007ti},
Albrecht-Skordis \cite{Albrecht:1999rm,Barnard:2007ta}, and Inverse Tracker
(or Inverse Power Law) \cite{PhysRevD.37.3406,Yashar:2008xx}
quintessence models for this work.
This is a diverse set of models each of which holds its own special
interest among researchers (see our discussion in the introductions of
\cite{Barnard:2007ta,Abrahamse:2007te,Bozek:2007ti,Yashar:2008xx} for
a brief review of the motivations and
\cite{Copeland:2006wr} to place these models in a more general
context).
One of our results is that these four quintessence models actually occupy very different
regions of the mode parameter space.  This tells us that the aspects
of $w(a)$ that are well measured by realistic experiments have the
potential to be extremely useful in discriminating among quintessence
models.  How much this potential is realized is of course related to the resolution
achieved by a given experiment in its mode parameter space, and that
issue is the subject of much of our analysis.

One of our key results  is that the goal of discriminating among
  these four models sets a concrete measure on the capabilities of future dark energy experiments.
  Experiments have to be somewhat better than DETF Stage 4 simulated
  experiments to fully meet this goal.

Sections \ref{sec2}, \ref{sec3}, and \ref{sec4}
describe our methods and Section \ref{sec5}
presents our detailed results,
while Section \ref{sec7} provides a discussion of the relevant statistical issues.
Our conclusion, Section \ref{sec6}, summarizes our results.

\section{Connection to our earlier work}
\label{sec2}

Our work builds very directly on our recent papers studying specific
quintessence models
\cite{Barnard:2007ta,Abrahamse:2007te,Bozek:2007ti,Yashar:2008xx}. We
refer the reader to those papers to learn more about our methods.  (An
appendix giving the technical details about our
  methods that are common to all these papers can be found in
  \cite{Abrahamse:2007te}.)  One product  of this earlier work
is a set of  Monte Carlo Markov Chains representing the distribution
of models that are consistent with specific Stage 2 simulated data
that is chosen to be consistent
with a cosmological constant cosmology.  Specifically, these chains represent
the distribution of possible scalar field parameters that are
consistent with a specific simulated Stage 2 data set.  Each
quintessence model has its own scalar field parameters and its own
chain representing the distribution in that space. Also, in each case
we base the Stage 2 data around a specific ``fiducial'' set of scalar field parameters
that are consistent with a cosmological constant at the $1-2 \sigma$
level.  Using these different fiducial models accounts in a rough way
for uncertainties in the outcomes of the Stage 2 experiments.

We use these chains for the work reported here by determining
the equation of state function $w(a)$ for each point on the chain and
projecting that function into the eigenmode-based space corresponding
to a particular simulated data set from Stage 3 or Stage 4 (as
discussed in detail below). When data from different scalar field
models is analyzed using the same eigenmodes, those modes
provide a common
parameter space in which scalar field models with different
``fundamental'' scalar field parameters can be compared on a common
footing.  In this usage the full chain represents how scalar field
models that are consistent with each
other at Stage 2 would be distributed in the eigenmode-based space
based on data from Stage 3 or 4.  Any discriminating power
among regions of the Stage 2 chains enabled by the higher stages represents progress over Stage 2 data, on
which the original chains are based.

\section{Generating Eigenmodes}
\label{sec3}

\begin{figure}[b]
\includegraphics[width=.5\textwidth]{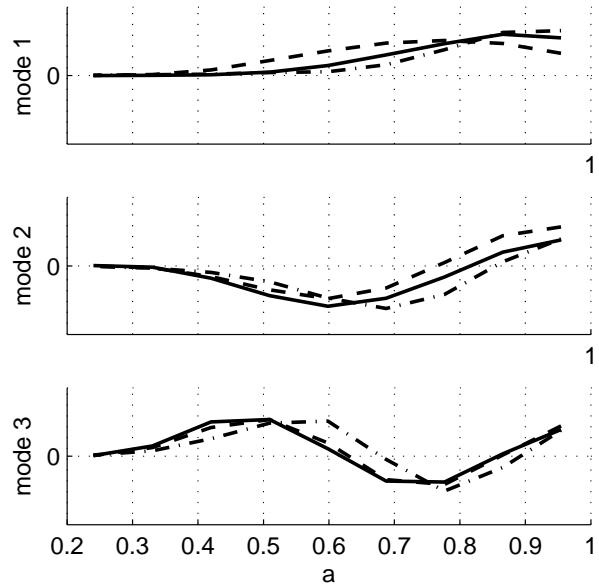}
\caption{
The first three best measured eigenmodes of $w(a)$.
The Stage 3 (dot-dash), Stage 4 ground (dash), and Stage 4 space (solid)
modes share a common general form at each mode level. Specific
differences can be related to various differences among the experiments
including how deeply a given experiment probes in redshift.
(Technically, these modes should each be represented by nine
discrete bin values.  The connecting lines guide the eye, and reflect
a likely ``continuum limit'' as discussed in \cite{Albrecht:2007qy}.)
}
\label{fig:Eig123}
\end{figure}

For this work, eigenmodes were generated using MCMC calculations.
We do this by breaking up the equation of state into nine bins linear in scale
factor from $a=.2$ to $a=1$ and using the value of $w+1$ in each of these bins
as parameters.
We then run Markov chains with these ``bin'' parameters in addition to cosmological parameters
in order to calculate a covariance matrix,
from which we extract the nine-by-nine covariance matrix for said ``bin'' parameters.
The eigenvectors of this matrix give us our eigenmodes,
while the eigenvalues are the squares of the uncorrelated error in each mode.
While these methods are slightly different from the Fisher matrix techniques of
\cite{Albrecht:2007qy}, the results are consistent, and our choice of
binning is driven by the analysis in \cite{Albrecht:2007qy}.
As with our previous work
\cite{Barnard:2007ta,Abrahamse:2007te,Bozek:2007ti,Yashar:2008xx},
 the simulated data sets are
constructed in a manner equivalent to the DETF simulated data. We do
not include cluster data (due to the technical difficulty of including
it discussed in \cite{Albrecht:2007qy}).
There are a number of possible considerations beyond the DETF
work (such as more carefully considering the impact of including cross
correlations among different photometric data types
\cite{Schneider:2006br,Zhan:2006gi}) that many
expect will lead to significant improvements over the DETF
projections.  For this paper we use the original DETF data models
for ease of comparison, except briefly in section \ref{sec5} where we
consider a simpleminded extension.

Plots of the first three eigenmodes are given in Figure
\ref{fig:Eig123} (ranked by how well each mode is measured).
All nine eigenmodes together form an orthonormal basis,
which is different for each data set.  The modes pick up additional
oscillations as one goes from best measured (mode 1) to less well
measured modes.

\section{Projection onto Eigenmodes}
\label{sec4}

\begin{figure}[b]
\includegraphics[width=.5\textwidth]{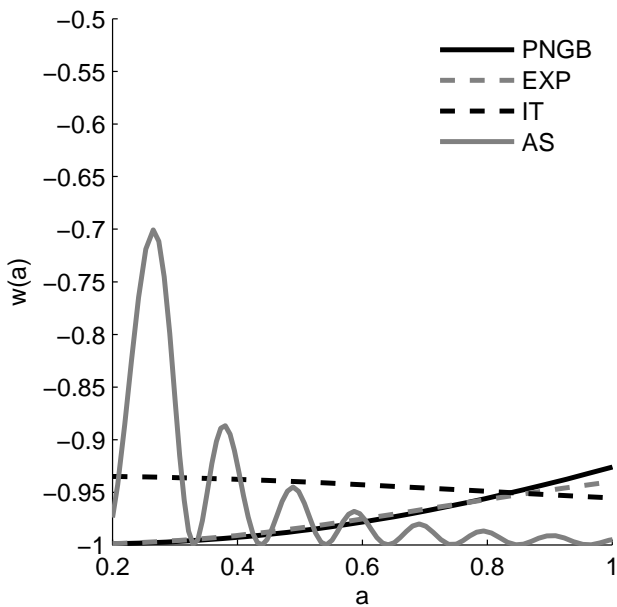}
\caption{
A characteristic $w(a)$ function for each of the four scalar field models considered in this paper.
}
\label{fig:4modelWofA}
\end{figure}

We  use the above eigenmodes to analyze four quintessence models of dark energy:
the Pseudo-Nambu-Goldstone Boson (PNGB),
Exponential, Albrecht-Skordis (AS), and Inverse Tracker (IT)
models.
Sample equation of state behavior of these models is illustrated in
Figure \ref{fig:4modelWofA}
\footnote{The current version of this paper corrects a technical error in the computation of the projection onto the eigenmodes vs earlier versions posted on the ArXiv and published in PRD.  These corrections affect the details of the distributions of the scalar field models in the mode plots, but our main points and results are unchanged, except for a somewhat greater importance attached to higher modes.}.
In each case, we will use points pulled from MCMC chains as representations of each model's parameter space.
The chains were run on DETF Stage 2 type data generated using a fiducial point in each model,
as in \cite{Barnard:2007ta,Abrahamse:2007te,Bozek:2007ti};
this gives us a fairly wide spread of parameter space for each model
and represents in a rough way the uncertainties in the outcomes of
the Stage 2 experiments.

\begin{figure}[t]
\includegraphics[width=0.5\textwidth]{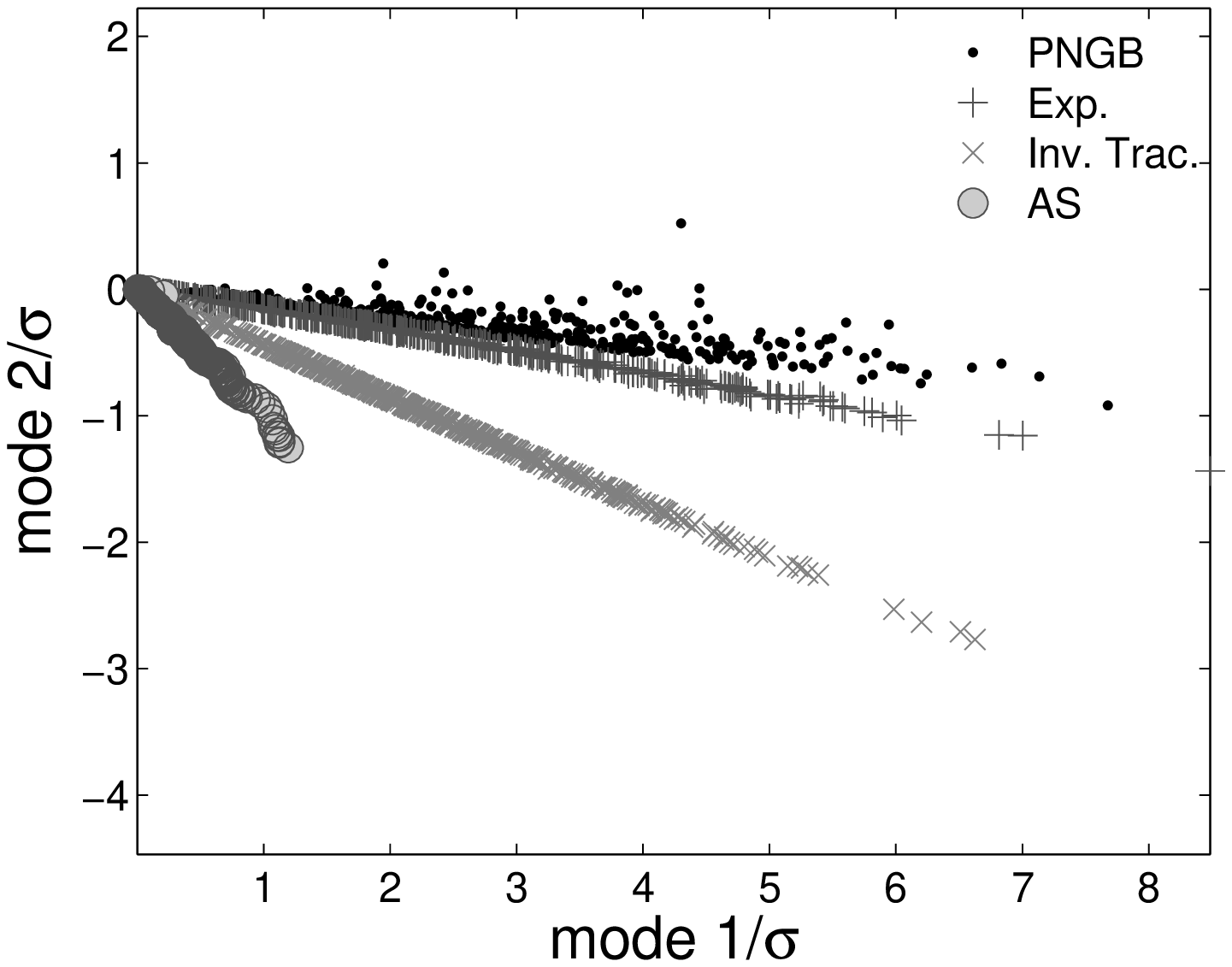}
\includegraphics[width=0.5\textwidth]{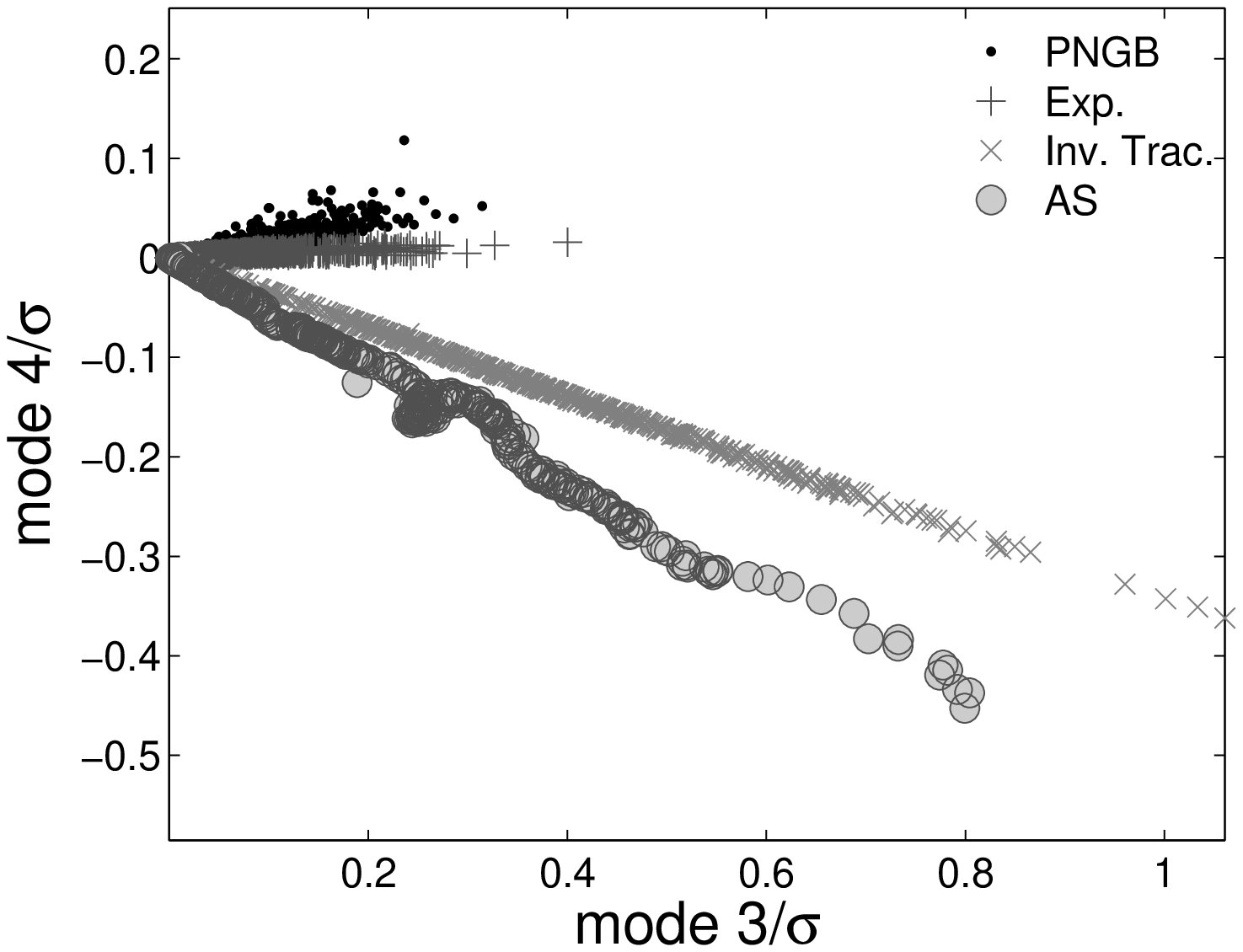}
\caption{{\bf Stage 3 photometric:}  A plot of the projections of the
PNGB, Exponential, AS, and Inverse Tracker models onto the first four Stage 3 photometric
eigenmodes.
The displayed points are from MCMC chains for each model,
and the scale of each axis is given by $\sigma_i$, the uncertainty in
measurements of that mode.
}
\label{fig:Mplot2DS3p}
\end{figure}

\begin{figure}[t]
\includegraphics[width=0.5\textwidth]{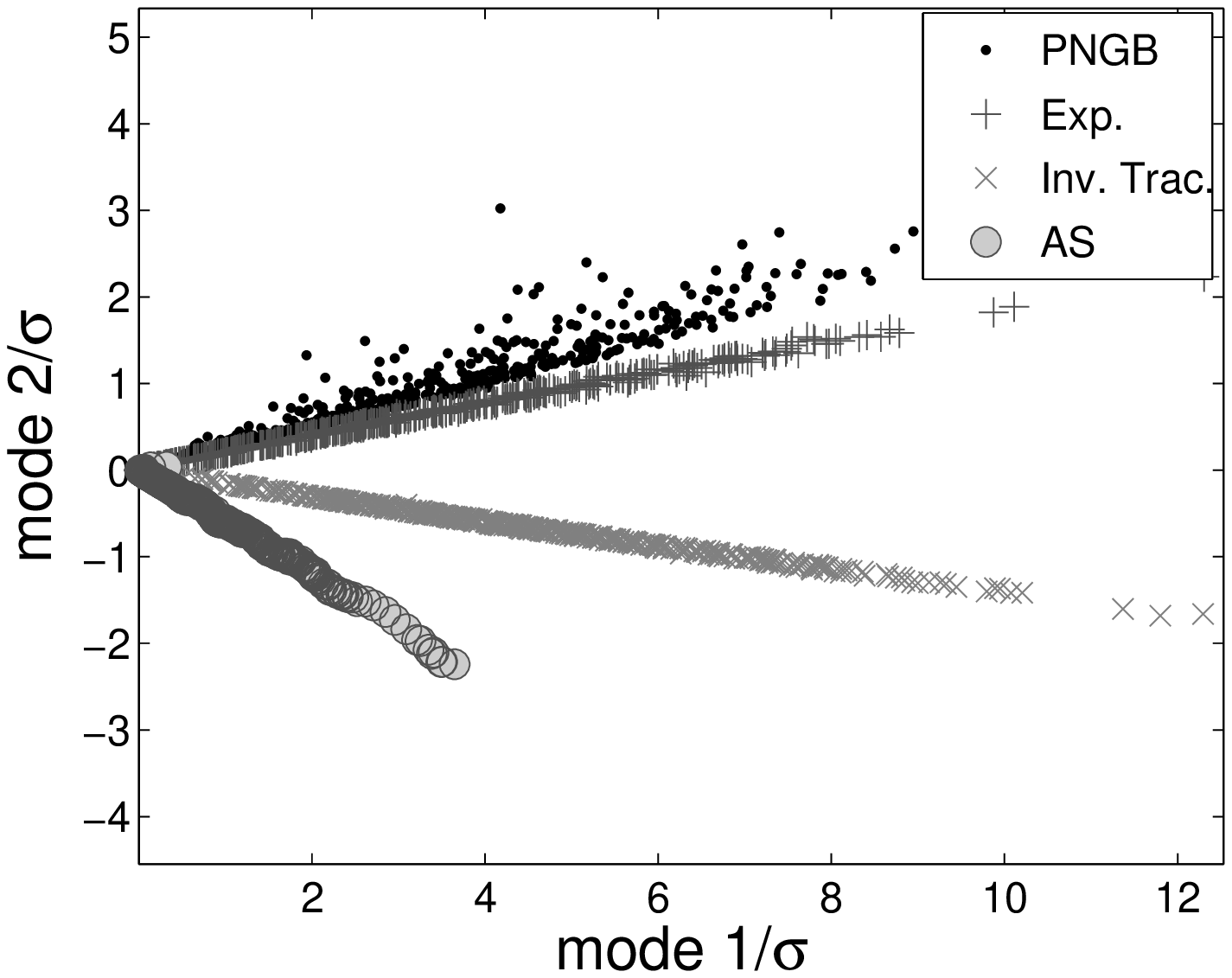}
\includegraphics[width=0.5\textwidth]{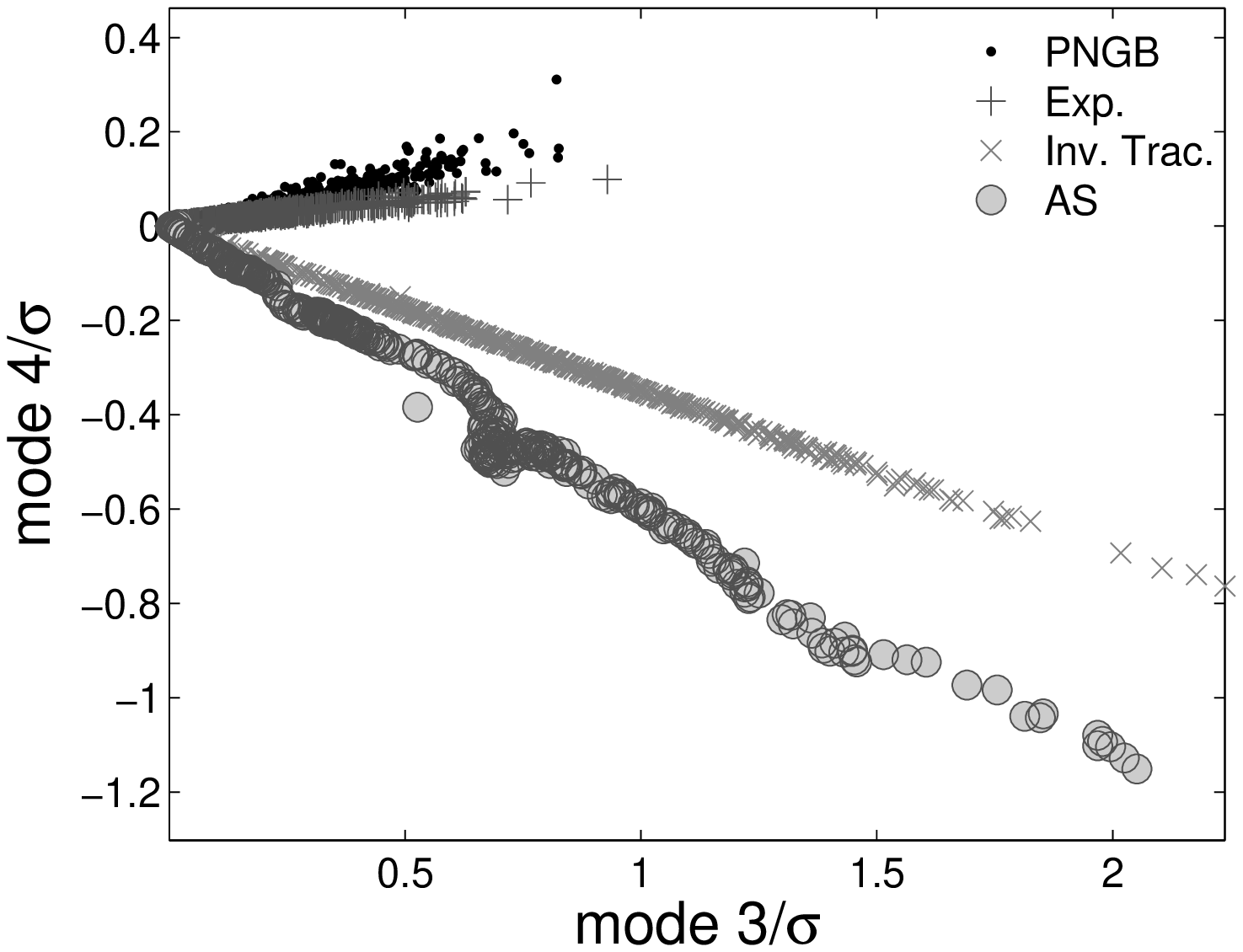}
\caption{{\bf Stage 4 ground:}  A plot of the projections of the
PNGB, Exponential, AS, and Inverse Tracker models onto the first four Stage 4 ground eigenmodes.
While the first mode is measured better by this data than by the Stage 3 data,
it is the improvement in the measurement of the second mode makes the models distinguishable
over a wide range of their parameters.
}
\label{fig:Mplot2DS4LST}
\end{figure}

\begin{figure}[t]
\includegraphics[width=0.5\textwidth]{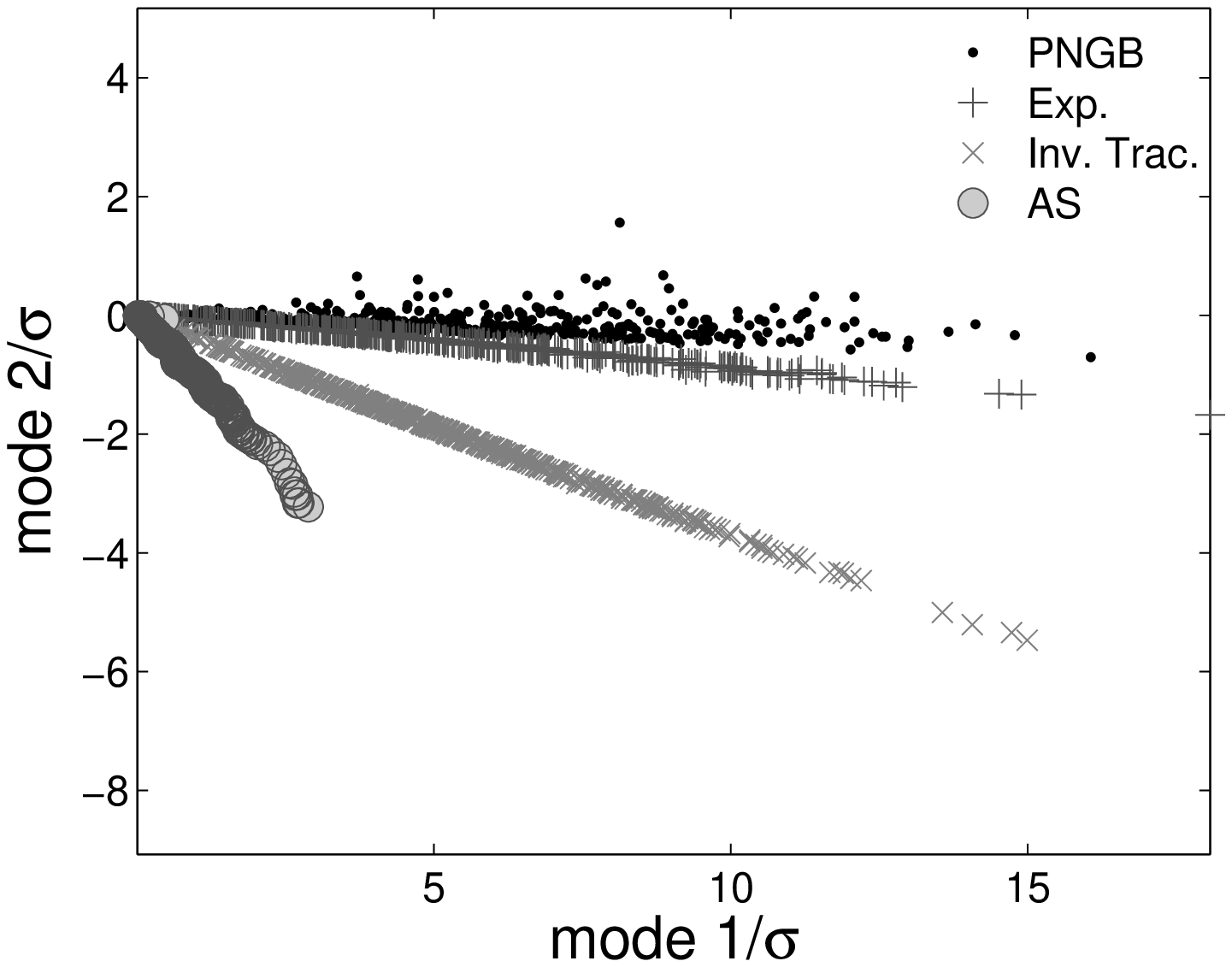}
\includegraphics[width=0.5\textwidth]{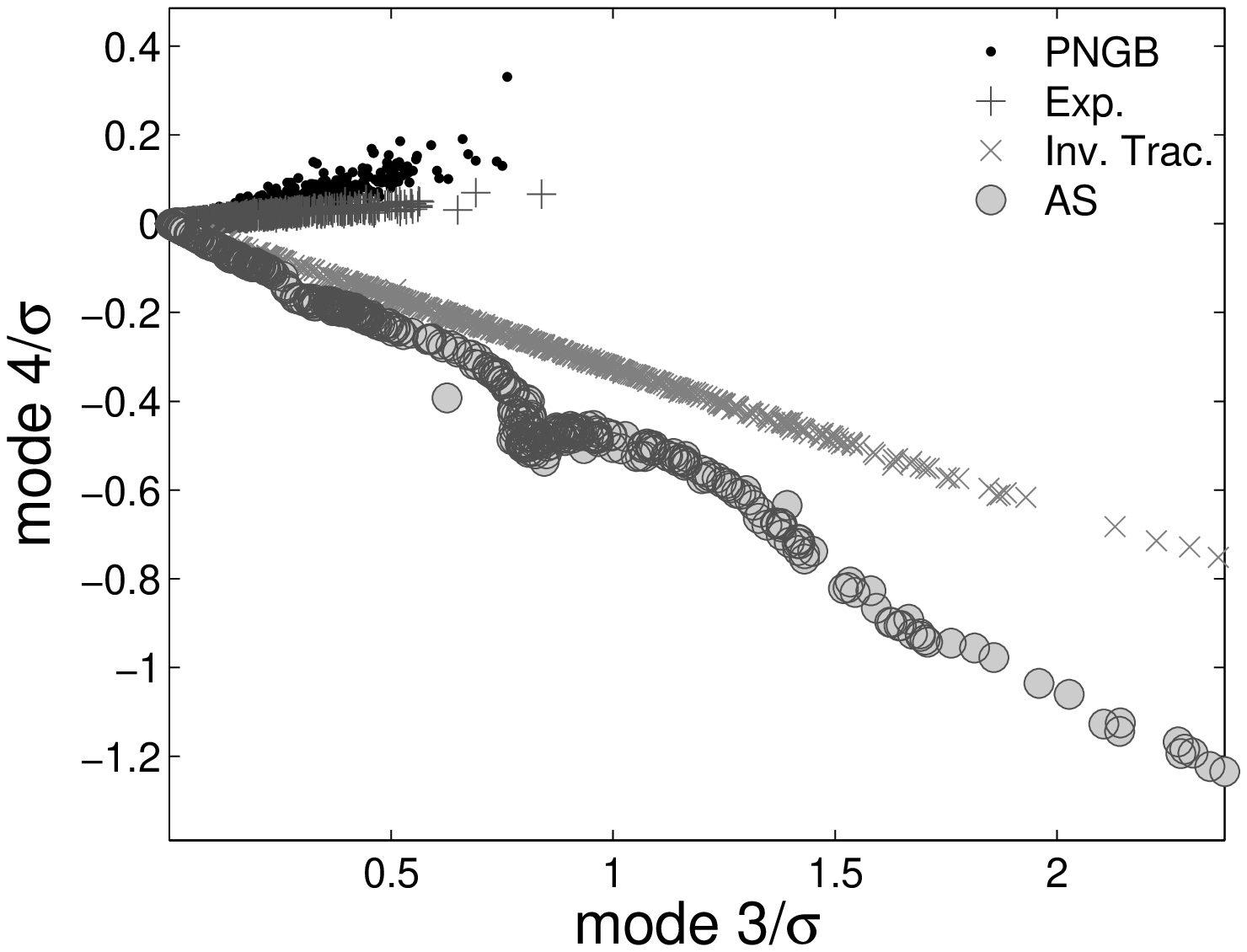}
\caption{{\bf Stage 4 space:}  A plot of the projections of the
PNGB, Exponential, AS, and Inverse Tracker models onto the first four Stage 4 space eigenmodes.
As with the Stage 4 ground plot, the most significant improvement,
in terms of model distinction, going from
the Stage 3 data to the Stage 4 space data is the measurement of the second mode.
}
\label{fig:Mplot2DS4sp}
\end{figure}

We use a simple algorithm to average the equation of state $w(a)$ in the nine scale factor bins, $w_j$.
These $w_j$ can be mapped into mode projections $m_i$ by matrix multiplication:
\begin{equation}
\sum_j E_{ij}(w_j+1)=m_i
\end{equation}
where $E_{ij}$ is the $j^{th}$ term of the $i^{th}$ eigenmode.
We use $w_j+1$ to center the eigenmodes around $w(a)=-1$,
so that $m_i=0$ for all $i$ is a cosmological constant cosmology.
The actual value of $m_i$ for a given $w(a)$ depends on the number of bins used,
so it is more convenient to look at $\frac{m_i}{\sigma_i}$,
where $\sigma_i$ is the square root of the $i^{th}$ covariance matrix eigenvalue.
This expresses the power in each eigenmode in units of its uncorrelated error,
and should be relatively stable as we change the number of modes by,
for example, refining the bin size.

When we consider how to display the range of quintessence models in the eigenmode space,
graphing the power in the first three modes in a
rotatable three dimensional plot can be fascinating.
However, as this does not lend itself to the static two dimensions of a paper,
it is more enlightening to view the modes two at a time
(an interested reader may contact the authors of this paper to view the three-dimensional versions).

Figure \ref{fig:Mplot2DS3p} shows Stage 2 distributions  of our four example
quintessence models represented in the first four modes of combined Stage 3
data (using photometric data and systematics designated
``optimistic'' by the DETF, just as was done in
\cite{Barnard:2007ta,Abrahamse:2007te,Bozek:2007ti,Yashar:2008xx}).
Because the axes of these plots are scaled in units of the error of each mode,
one can get an intuitive idea of the Stage 3 resolving power by noting
that Stage 3 data should roughly resolve areas of unit size in these
plots.

It is interesting to note how the models examined here occupy very
distinct portions of the ``mode space'' except for where they meet at the origin
($w(a)=-1$).
We should mention here that these figures in many cases display only a subset
of the total space accessible by the models,
because the parameter spaces of the models were restricted in the MCMC chains.
This is particularly true of the AS model, for which we expect that the kinks and wiggles in the displayed
distribution will characterize a broader distribution of such
features.  This may also be an issue for the PNGB model,
which was restricted in the MCMC analysis to the concave down portion of its potential.
The Exponential model, however, has a concave up potential
and gives us some idea of where the concave up portion of the PNGB model may
lie in the mode space. As discussed in
\cite{Barnard:2007ta,Abrahamse:2007te,Bozek:2007ti,Yashar:2008xx},
most of these restrictions were required to eliminate parameters that would be
completely unconstrained by even the best future data.

On a similar note, there is a small fraction of the points ($<1\%$) in the Inverse Tracker model
that never display tracking behavior,
but instead display the thawing \cite{Caldwell:2005tm}
behavior of the PNGB and Exponential models;
this can be seen in the plots as a handful of outlying points above the main
concentration of Inverse Tracker model points.

It is of interest that, as we look at parameterizations of increasing
distinction from the cosmological constant,
we see a consistent increase in the amplitude of their first mode projections.
As such, we might look at the first mode projection as a signal
of the presence of one of these scalar field models,
but only in extreme cases will it help us distinguish between them
(noting that a very large value would actually rule out the AS model).
It is clear that it is the higher modes that will distinguish between models,
though the plot shows that Stage 3 data will not do this very well.

The authors of \cite{Huterer:2006mv} also made plots using similar modes to ours,
but their approach differs.
In \cite{Huterer:2006mv} the potentials are drawn from an abstract
continuous space that covers large ranges of possibilities, whereas our
potentials are drawn from specific scalar field models which each only
have certain classes of behaviors even as the parameters are varied
fully.
These differences have allowed us to discover a much more
striking structure than can be seen in \cite{Huterer:2006mv}.

In Figure \ref{fig:Mplot2DS4LST} and \ref{fig:Mplot2DS4sp} we show the
mode projection plots for Stage 4 space and ground data sets,
respectively (using the same DETF ``optimistic'' combinations used in
\cite{Barnard:2007ta,Abrahamse:2007te,Bozek:2007ti,Yashar:2008xx}).
We can see from these plots that the overall shape of the distribution of
projected models remains very similar in the first two modes.
The Stage 4 data sets do a better job of measuring the first mode than Stage 3,
a much better job of measuring the second mode,
and even begin to give resolution of the third mode that is of
marginal value in resolving these models.
As a result, for a scenario in which the Stage 3 projects detected the presence
of quintessence at the level of a few sigma,
the Stage 4 data should be able to discriminate between
this sample of models to at least that level.

\section{Resolving the quintessence models}
\label{sec5}

\begin{figure}[t]
\includegraphics[width=.5\textwidth]{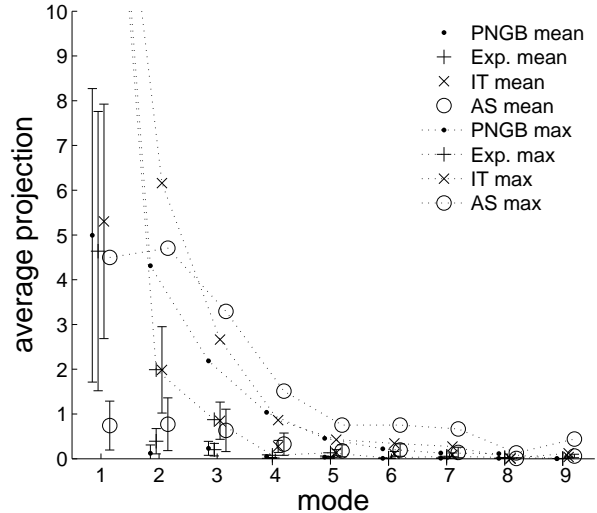}
\caption{A plot of the mean (with bars at one standard deviation)
and maximum projections of the examined models into the Stage 4 space mode-space, in terms of the mode uncertainties.
It can be seen that for nearly all of the model points,
the projections onto the modes higher than three are negligible.
Though the some models have outlying points that have some power in the higher modes,
the bulk of each model does not.
}
\label{fig:MeanProjS4sp}
\end{figure}

\begin{figure}[t]
\includegraphics[width=0.5\textwidth]{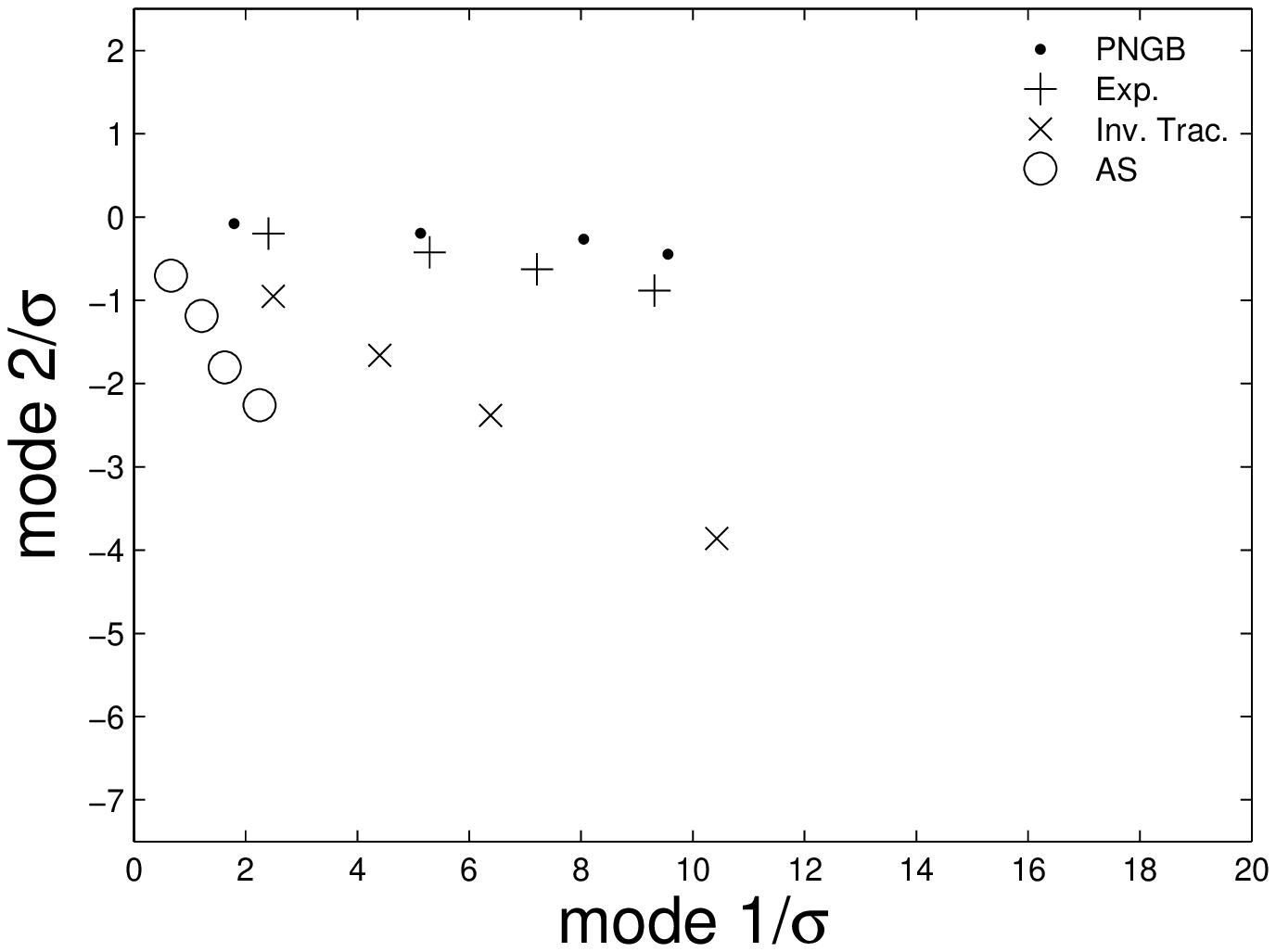}
\includegraphics[width=0.5\textwidth]{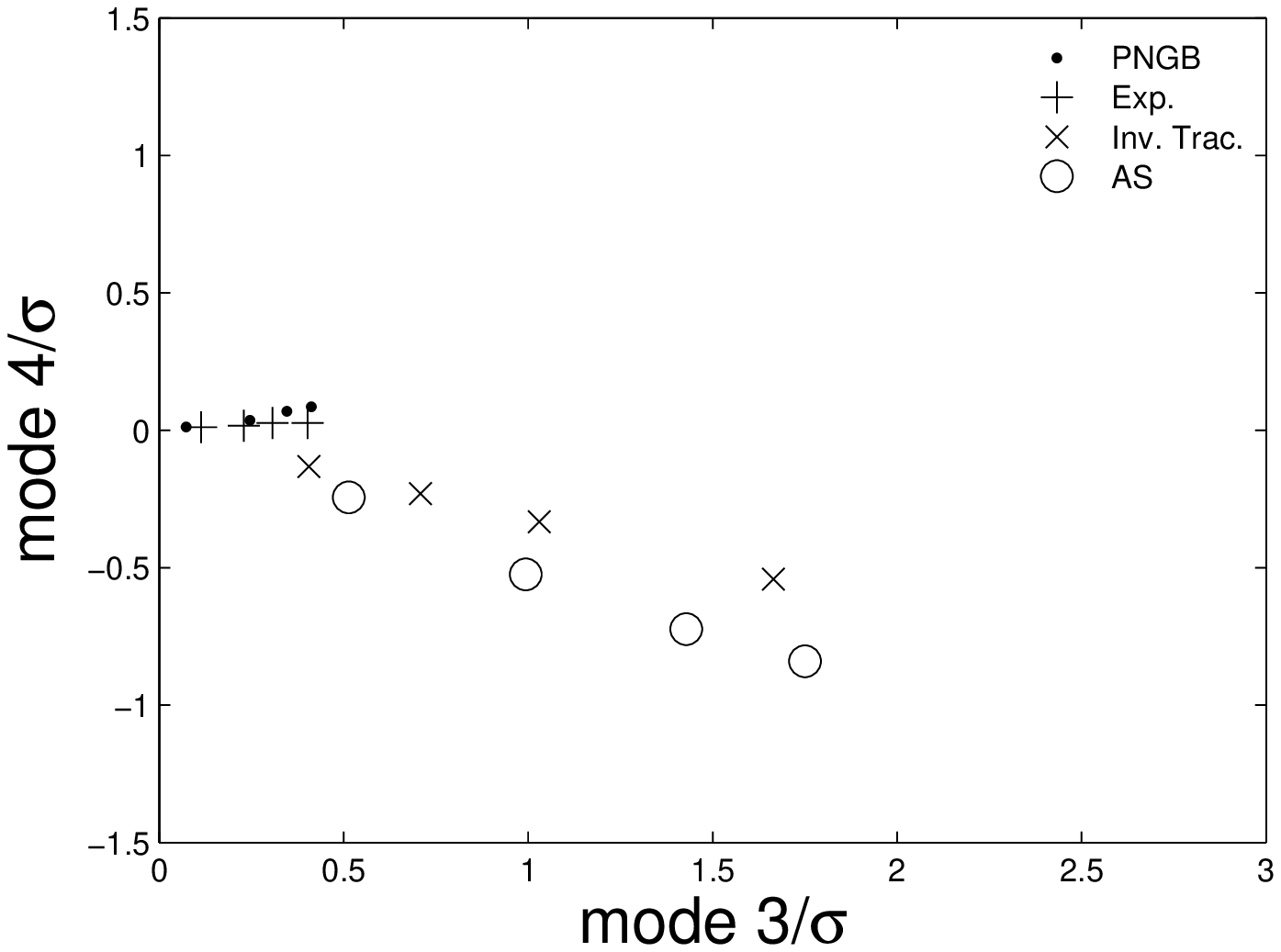}
\caption{The four test points for each model (16 test points in all) represent possible
  experimental outcomes. The test points are plotted here in the
  eigenmodes for Stage 4 space. One can compare this figure with
  Fig. \ref{fig:Mplot2DS4sp} to see the degree to which the full range
  of behavior of each model is represented by the test points.}
\label{fig:Mplot2DS4spTP}
\end{figure}

With the models projected onto the eigenmode space,
we then have a common space in which to compare the different scalar
field models and evaluate how well coming experiments will distinguish
among them.
We can use mode uncertainties as a metric to find how measurably different each quintessence model will be.
Informed by the plots of mode projections,
we can reasonably expect that the first mode will dominate this measure,
but the second and third modes will play a part in
setting the minimum separation.
As for the higher modes, we can expect to see these swamped by their uncertainty;
a combination of the poor measurement and the lack of power in these modes
by our quintessence models makes them largely irrelevant.
Figure \ref{fig:MeanProjS4sp} gives information about the distribution
of mode amplitudes for our chains. From Fig. \ref{fig:MeanProjS4sp}
one can see that the measurement error in the higher
modes will be much larger than the distribution of physically interesting values.
This gives us, in effect, a prior on the value of these modes that is much stronger than
the data constraints from even Stage 4 experiments.
Therefore, this calculation will actually be done using only the first three eigenmodes.

\begin{table}[h]
\centering
\caption{Stage 3 photometric.  This is a table of the minimum $\chi^2$ (ignoring the smallest $1\%$ of $\chi^2$ values). The $99\%$ confidence level is a $\chi^2$ of $11.36$.
Notably, the only comparisons that rise to that level are between
the test points with large first mode projections and the AS model,
telling us that this is mostly a first mode measurement.
As discussed in the text, this table gives the values of $\chi^2$
where the curves in Fig. \ref{fig:EigSepS3p} sharply approach the $x$-axis.
}
\begin{tabular}{| l r| c | c | c | c |}
	\hline
\hline
PNGB & &PNGB&Exp.&IT&AS \\
\hline
point 1& &   0.001 &  0.001 &  0.1 &  0.3 \\
point 2& &   0.002 &  0.01 &  0.6 &  2.9 \\
point 3& &   0.004 &  0.04 &  1.4 &  8.5 \\
point 4& &   0.01 &  0.04 &  1.8 &  12.7 \\
\hline
Exp. & &PNGB&Exp.&IT&AS \\
\hline
point 1& &   0.004 &  0.001 &  0.1 &  0.5 \\
point 2& &   0.01 &  0.001 &  0.5 &  2.8 \\
point 3& &   0.03 &  0.001 &  0.8 &  6.0 \\
point 4& &   0.1 &  0.01 &  1.2 &  11.3 \\
\hline
IT & &PNGB&Exp.&IT&AS \\
\hline
point 1& &   0.1 &  0.1 &  0.001 &  0.3 \\
point 2& &   0.4 &  0.3 &  0.0003 &  1.0 \\
point 3& &   0.8 &  0.6 &  0.001 &  3.2 \\
point 4& &   2.3 &  1.6 &  0.01 & 13.5 \\
\hline
AS & &PNGB&Exp.&IT&AS \\
\hline
point 1& &   0.1 &  0.1 &  0.05 &  0.0001 \\
point 2& &   0.3 &  0.3 &  0.1 &  0.001 \\
point 3& &   0.6 &  0.6 &  0.3 &  0.001 \\
point 4& &   1.0 &  1.0 &  0.4 &  0.01 \\
	\hline
\end{tabular}
\label{tab:Table1}
\end{table}

\begin{table}[h]
\centering
\caption{Stage 4 Ground.
This is a table of the minimum $\chi^2$ (ignoring the smallest $1\%$ of $\chi^2$ values).
Again, the low first mode projections for the AS model are responsible
for the highest $\chi^2$ values,
but we can also see significant separation for test points that
have first mode projections in the range of the model they are being
compared to.
As discussed in the text, this table gives the values of $\chi^2$
where the curves in Fig. \ref{fig:EigSepS4LST} sharply approach the $x$-axis.}
\begin{tabular}{| l r | c | c | c | c |}
	\hline
\hline
PNGB & &PNGB&Exp.&IT&AS \\
\hline
point 1& &   0.003 &  0.005 &  0.2 &  0.7 \\
point 2& &   0.004 &  0.05 &  1.9 &  6.6 \\
point 3& &   0.01 &  0.2 &  4.8 & 16.8 \\
point 4& &   0.03 &  0.1 &  6.3 & 23.5 \\
\hline
Exp. & &PNGB&Exp.&IT&AS \\
\hline
point 1& &   0.01 &  0.001 &  0.3 &  1.3 \\
point 2& &   0.04 &  0.002 &  1.6 &  6.8 \\
point 3& &   0.1 &  0.003 &  2.9 & 12.7 \\
point 4& &   0.3 &  0.01 &  4.5 & 21.5 \\
\hline
IT &  &PNGB&Exp.&IT&AS \\
\hline
point 1& &   0.7 &  0.6 &  0.001 &  0.9 \\
point 2& &   2.2 &  1.7 &  0.001 &  3.0 \\
point 3& &   4.6 &  3.5 &  0.002 &  7.1 \\
point 4& &  12.9 & 8.9 &  0.03 & 33.1 \\
\hline
AS & &PNGB&Exp.&IT&AS \\
\hline
point 1& &   0.7 &  0.6 &  0.2 &  0.001 \\
point 2& &   2.0 &  1.8 &  0.7 &  0.002 \\
point 3& &   4.2 &  4.0 &  1.6 &  0.004 \\
point 4& &   6.7 &  6.2 &  2.3 &  0.04 \\
	\hline
\end{tabular}
\label{tab:Table2}
\end{table}

\begin{table}[h]
\centering
\caption{Stage 4 space.
This is a table of the minimum $\chi^2$ (ignoring the smallest $1\%$ of $\chi^2$ values).
Again, the low first mode projections for the AS model are responsible
for the highest $\chi^2$ values,
but we can also see significant separation for test points that
have first mode projections in the range of the model they are being compared to.
As discussed in the text, this table gives the values of $\chi^2$
where the curves in Fig. \ref{fig:EigSepS4sp} sharply approach the $x$-axis.}
\begin{tabular}{| l r | c | c | c | c |}
	\hline
\hline
PNGB & &PNGB&Exp.&IT&AS \\
\hline
point 1& &   0.01 &  0.01 &  0.4 &  1.7 \\
point 2& &   0.01 &  0.05 &  2.9 & 15.6 \\
point 3& &   0.02 &  0.2 &  7.2 & 39.6 \\
point 4& &   0.04 &  0.2 & 9.3 & 57.6 \\
\hline
Exp. & &PNGB&Exp.&IT&AS \\
\hline
point 1& &   0.02 &  0.002 &  0.5 &  3.1 \\
point 2& &   0.05 &  0.003 &  2.4 & 15.7 \\
point 3& &   0.1 &  0.01 &  4.2 & 29.6 \\
point 4& &   0.3 &  0.02 &  6.7 & 52.5 \\
\hline
IT & &PNGB&Exp.&IT&AS \\
\hline
point 1& &   0.8 &  0.7 &  0.003 &  2.0 \\
point 2& &   2.3 &  1.9 &  0.002 &  6.5 \\
point 3& &   4.9 &  3.9 &  0.002 &  16.4 \\
point 4& &  13.3 & 9.9 &  0.05 & 65.1 \\
\hline
AS & &PNGB&Exp.&IT&AS \\
\hline
point 1& &   0.7 &  0.7 &  0.3 &  0.001 \\
point 2& &   2.2 &  2.0 &  1.0 &  0.01  \\
point 3& &  4.8 & 4.6 &  2.4 &  0.01 \\
point 4& &  7.3 & 7.0 & 3.4 &  0.01 \\
	\hline
\end{tabular}
\label{tab:Table3}
\end{table}

\begin{table}[h]
\centering
\caption{Stage 4 space, with experimental uncertainty reduced by $2/3$
  in each mode.
This is a table of the minimum $\chi^2$ (ignoring the smallest $1\%$ of $\chi^2$ values).
}
\begin{tabular}{| l r | c | c | c | c |}
	\hline
\hline
PNGB & &PNGB&Exp.&IT&AS \\
\hline
point 1& &   0.01 &  0.01 &  0.8 &  3.9 \\
point 2& &   0.01 &  0.1 &  6.4 & 35.0 \\
point 3& &   0.04 &  0.4 &  16.1 & 89.0 \\
point 4& &   0.1 &  0.4 & 21.0 & 129.6 \\
\hline
Exp. & &PNGB&Exp.&IT&AS \\
\hline
point 1& &   0.04 &  0.01 &  1.2 &  6.9 \\
point 2& &   0.1 &  0.01 &  5.5 & 35.4 \\
point 3& &   0.3 &  0.01 &  9.6 & 66.5 \\
point 4& &   0.7 &  0.05 &  15.1 & 118.1 \\
\hline
IT & &PNGB&Exp.&IT&AS \\
\hline
point 1& &   1.8 &  1.5 &  0.01 &  4.5 \\
point 2& &   5.3 &  4.3 &  0.003 &  14.5 \\
point 3& &   11.0 &  8.7 &  0.01 &  36.9 \\
point 4& &  30.0 & 22.3 &  0.1 & 146.5 \\
\hline
AS & &PNGB&Exp.&IT&AS \\
\hline
point 1& &   1.6 &  1.4 &  0.7 &  0.002 \\
point 2& &   4.9 &  4.6 &  2.3 &  0.01  \\
point 3& &  10.9 & 10.4 &  5.5 &  0.01 \\
point 4& &  16.5 & 15.7 & 7.7 &  0.1 \\
	\hline
\end{tabular}
\label{tab:Table4}
\end{table}

Assuming Stage 2 data as discussed above, higher stage data might come
from a universe based on any of the quintessence models represented by
points in
our mode space plots. Our next step is to scan all possible data outcomes
and evaluate the potential discriminating power. We represent the
Stage 2 distribution of each quintessence model, by choosing four
``test points'' spread evenly across the mode space.  Four test points
each for four quintessence models gives a total of sixteen ``data
outcomes'' which are meant to represent (in a rough way) the full range of
possibilities. The distribution of our test points in mode space is
shown in Fig. \ref{fig:Mplot2DS4spTP}.
These ``test points" were chosen by their first mode projections,
which places them in sequence along the very nearly linear regions the
models cover \footnote{The fact that each model occupies a somewhat
  ``one dimensional'' region in the mode space is part of why it seems
  reasonable to represent the distribution of data outcomes by the
  four test points for each model.}.
For each of the sixteen test points, we analyze the ability of data
centered on the test point to exclude other points on the chains.
Note that in Fig. \ref{fig:Mplot2DS4spTP} the test points are not
represented with noise, which is expected in any data
set (at a level given by the $\sigma_i's$) and which is reflected in
our quantitative analysis.

\begin{figure*}[p]
{\includegraphics[width=\textwidth]{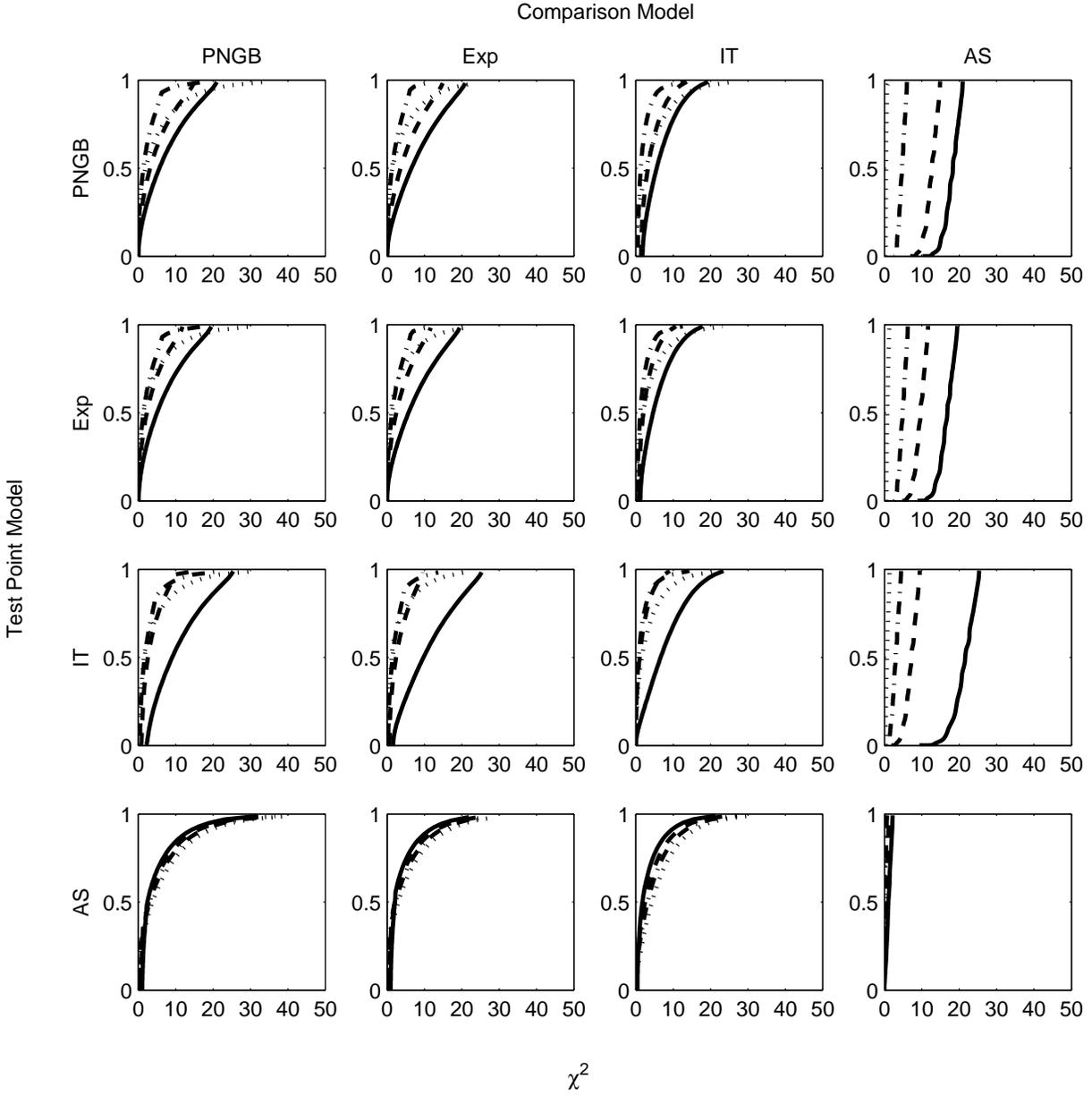}}
\caption{{\bf Stage 3 Photometric}.  These plots show the
  distributions of $\chi^2$ as each test point is compared with each
  comparison model. Specifically we plot
the fraction of $\chi^2$ values less than $\chi^2$ given on the
  $x$-axis.   The rows of this figure correspond to which model the test points have been pulled from
(as seen in Figure \ref{fig:Mplot2DS4spTP}), while the columns
  correspond to the model which is compared to those test points.
In each plot, the test point closest to the origin of the mode space (and thus, a $\Lambda$CDM model)
is denoted by a dotted line, the next closest as a dot-dashed line, the next as a dashed line,
and the farthest as a solid line.
The $99\%$ confidence interval for three parameters is
$\chi^2=11.36$.  The relatively sharp left-hand edges of these curves
are an interesting feature (related to the gaps between models in mode
  space) which is discussed in the text and Table \ref{tab:Table1}
}
\label{fig:EigSepS3p}
\end{figure*}

\begin{figure*}[p]
{\includegraphics[width=\textwidth]{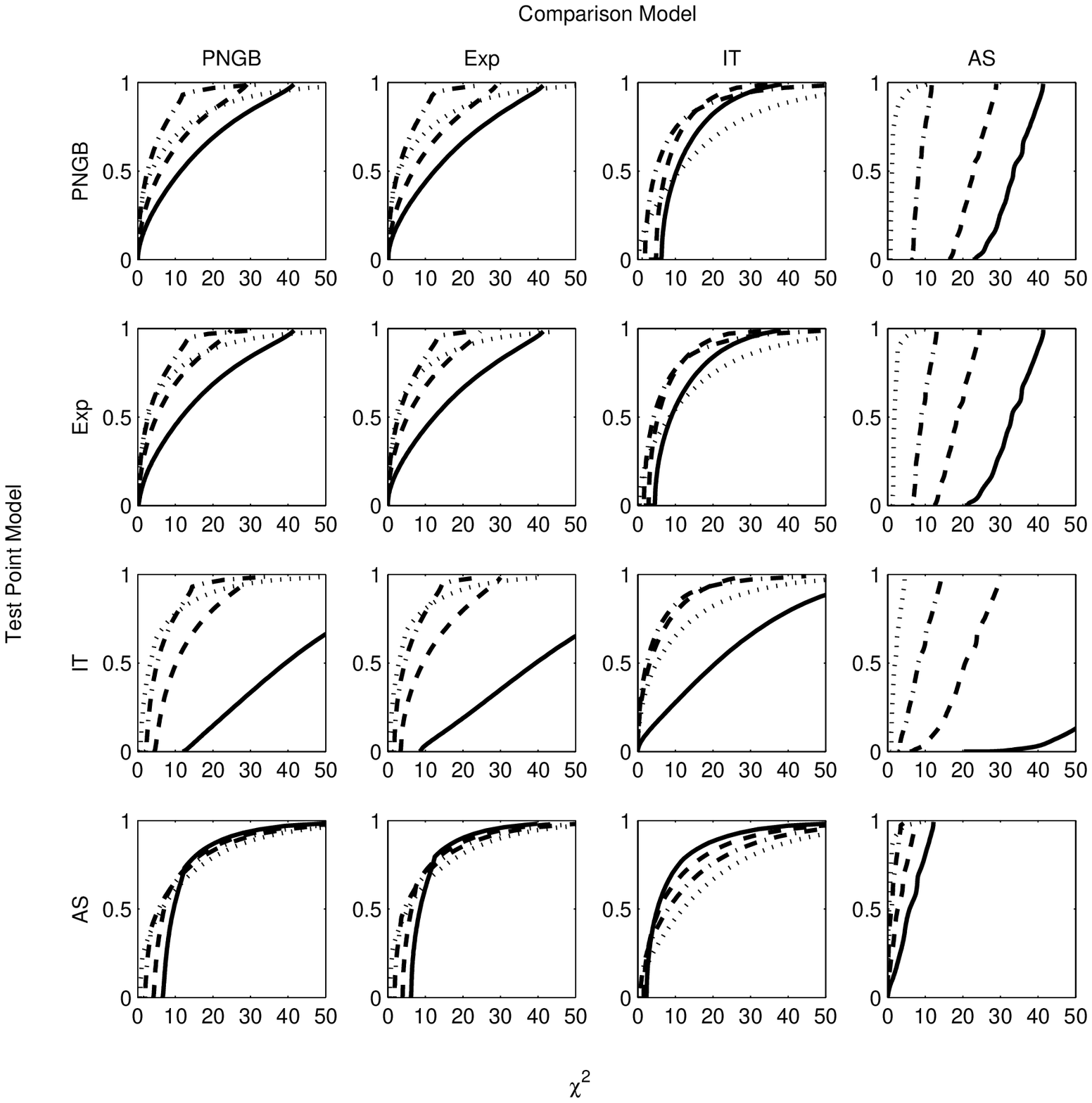}}
\caption{{\bf Stage 4 Ground}.   These plots show the
  distributions of $\chi^2$ as each test point is compared with each
  comparison model. Specifically we plot
the fraction of $\chi^2$ values less than $\chi^2$ given on the
  $x$-axis.   The rows of this figure correspond to which model the test points have been pulled from
(as seen in Figure \ref{fig:Mplot2DS4spTP}), while the columns
  correspond to the model which is compared to those test points.
In each plot, the test point closest to the origin of the mode space
is denoted by a dotted line, the next closest as a dot-dashed line, the next as a dashed line,
and the farthest as a solid line.
The $99\%$ confidence interval for three parameters is
$\chi^2=11.36$. The relatively sharp left-hand edges of these curves
are an interesting feature (related to the gaps between models in mode
  space) which is discussed in the text and Table \ref{tab:Table2}
}
\label{fig:EigSepS4LST}
\end{figure*}

\begin{figure*}[p]
{\includegraphics[width=\textwidth]{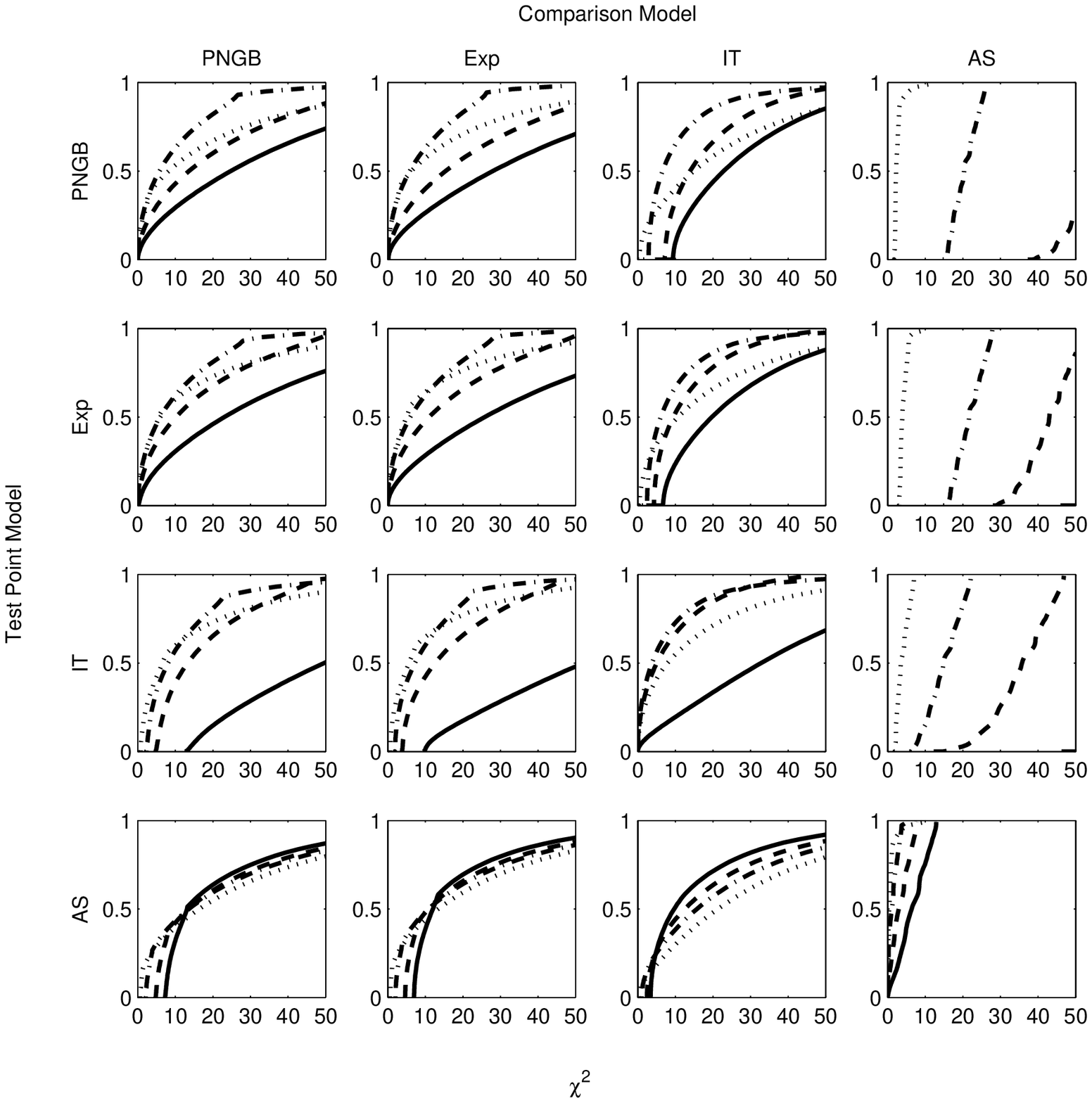}}
\caption{{\bf Stage 4 Space}.    These plots show the
  distributions of $\chi^2$ as each test point is compared with each
  comparison model. Specifically we plot
the fraction of $\chi^2$ values less than $\chi^2$ given on the
  $x$-axis.   The rows of this figure correspond to which model the test points have been pulled from
(as seen in Figure \ref{fig:Mplot2DS4spTP}), while the columns
  correspond to the model which is compared to those test points.
In each plot, the test point closest to the origin of the mode space
is denoted by a dotted line, the next closest as a dot-dashed line, the next as a dashed line,
and the farthest as a solid line.
The $99\%$ confidence interval for three parameters is
$\chi^2=11.36$.  The relatively sharp left-hand edges of these curves
are an interesting feature (related to the gaps between models in mode
  space) which is discussed in the text and Table \ref{tab:Table3}
}
\label{fig:EigSepS4sp}
\end{figure*}

The curves in Figs. \ref{fig:EigSepS3p}, \ref{fig:EigSepS4LST}, and
\ref{fig:EigSepS4sp} show the fraction of model points with  $\chi^2$
less than a specific value, given on the $x$-axis\footnote{ A
  technical clarification: The ``model points'' used here are the
  unique points taken from
MCMC chains, so the repeated points in the original chain, where the
algorithm
did not step to another point, are counted only once for the purposes of these plots.
As discussed in section \ref{sec7}, we expect this omission to impact
the detailed shapes of the curves shown here but not the minimum
$\chi^2$ value (where the curve hits the $x$-axis). Since we focus on
this  minimum $\chi^2$ value in our analysis we do not consider the
effect this has on the graphs significant or meaningful enough for
our analysis to justify the increased computation including the repeated points would entail.}.
To read these plots, start with the labels on the left hand side.
These read PNGB (Pseudo-Nambu-Goldstone Boson), EXP (Exponential), IT (Inverse Tracker), and AS (Albrecht-Skordis),
and label which model the ``test points'' used in that row of plots were pulled from.
Then look to the top of the figure, where the same model labels mark the columns;
these label which model the test points are being compared to in each column.
In each plot, four functions are graphed, $\chi^2$ vs. the fraction of the compared model's points
that have that $\chi^2$ or lower relative to the test point in question.
For example, looking at the PNGB vs. PNGB plot, the curves represent
how far the rest of the PNGB model is from each PNGB test point.
The point where each function touches the x-axis (the minimum $\chi^2$)
is in this case a loose measure of how densely the MCMC chain populates
the mode space at that test point.
However, if one were to look at the PNGB row and the AS column,
the graph there represents how far the PNGB test points are from all of the points in the
AS chain, and the minimum $\chi^2$ of each function represents by what $\chi^2$ each of the
PNGB test points would rule out the entire AS model.

The sharp left-hand edges of curves in Figs. \ref{fig:EigSepS3p},
\ref{fig:EigSepS4LST}, and \ref{fig:EigSepS4sp} imply very strong
discriminating power at the level of $\chi^2$ given by the point on
the $x$-axis where the edge is located.  We have organized information
about this important feature in the following tables:
Tables \ref{tab:Table1}, \ref{tab:Table2}, and \ref{tab:Table3}
give numerical results
for the minimum $\chi^2$ one could estimate
from Figures \ref{fig:EigSepS3p}, \ref{fig:EigSepS4LST}, and \ref{fig:EigSepS4sp}.
In general (due to the outlying points and other factors), one expects
some low level tails even on the otherwise sharp rising edges.  In
order to make sure we are quantifying the true edge of the figure we
ignore the closest $1\%$ of points from the compared model.  This is
equivalent to finding the $\chi^2$ where each plot crosses the $.01$
fraction mark.
For completeness, we have again included the Exponential model,
though we do not expect any experiment to distinguish between it and the PNGB model.
The numbers reported here reinforce the intuition we gain from the mode plots:
only in extreme cases will Stage 3 distinguish between the models.
The only examples that rise above the level of $99\%$ confidence
(which is a $\chi^2$ of $11.36$ for the three parameters used here)
are due to the previously mentioned observation that the AS model has
a much smaller range of amplitudes in the first mode
than the other three models.

The Stage 4 ground and Stage 4 space $\chi^2$ values show significant improvement over Stage 3.
Again, the largest $\chi^2$ values come from the large first mode separation between the last test point
for each model and the nearest AS model point.

Looking at Tables \ref{tab:Table2} and
\ref{tab:Table3} it appears that the Stage 4 data lie on some kind of
threshold:  There are quite a few entries greater than $11.36$
 (indicating exclusion at $99$\% confidence) but plenty that are lower.
To explore the nature of this threshold a bit more, Table
\ref{tab:Table4} presents
$\chi^2$ values for a hypothetical data set that would improve the
$\sigma_i$ by a factor of $1.5$ over Stage 4 space. Due to a variety of
considerations (including those discussed in
\cite{Schneider:2006br,Zhan:2006gi}) many believe such improvements (or
even much better ones) over the DETF modeling of Stage 4 to be
realistic for some experiments.  In Table \ref{tab:Table4} there are
indeed a great many more entries greater than $11.36$, further
supporting the notion of a threshold around Stage 4.
As a reference point, the $\sigma_i$ of the Stage 4 data sets are a
factor of about $4$ or $5$ smaller than the Stage 3 $\sigma_i$ for
most modes.  To get a more complete understanding of which experiments
might achieve (or beat) the level of model discrimination shown in
Table \ref{tab:Table4} one would have to undertake detailed modeling
of alternative experiments and data analysis schemes. Such a
systematic analysis is not the subject of this paper, and we leave
that for future work.

\section{Discussion of Statistics}
\label{sec7}

The MCMC chains of the selected scalar field models in this analysis
are used as a means to represent the phenomenology associated with each
model, that is, to represent the full spread of points in mode space the
model can occupy.
This is not to be confused with the set of priors that would be used in
a Bayesian analysis.
The question we seek to answer with Figs.
\ref{fig:EigSepS3p}, \ref{fig:EigSepS4LST}, and \ref{fig:EigSepS4sp},
and Tables \ref{tab:Table1} to \ref{tab:Table4},
is ``If the universe is a single realization of a particular model,
how well will a given experiment rule out other
models/phenomenologies?''
The method we use to answer this is to take the probability of observing
the observables generated by that realization
(for a given experiment) at all other points in the eigenmode space,
and then find the highest of those probabilities (smallest $\chi^2$)
out of our sets of model-based phenomenologies.
We can then speak to the power of the experiments to distinguish between
models.

Because we are only interested in the regions of closest approach
between
a model and a hypothetical realization, the exact distribution of the
points
in the chains used for this work is largely unimportant,
so long as it does a reasonable job of exploring the possibilities of
each.  Our focus on the regions of closest approach is really a luxury
which is enabled by the remarkable structure we have found in the
eigenmode space.  An outcome with more overlap (such as we do see
between the Exponential and PNGB models) might draw one into a Bayesian
analysis where priors on the different model distributions in mode space
would play a critical role.  One can loosely think of our approach as a
very conservative one that allows for very unprincipled theorists who
might cook up reasons to place delta function priors on the point of
closest approach.  We have shown that high quality Stage 4 experiments
could rule out entire models, even under these adverse circumstances.

There is an important assumption built into our analysis. Our choice of
four test points spread out along the mode curves for each model implies
that we are giving "equal weight" to the actual Universe residing
anywhere along the mode space curves.  In the current confused state of
dark energy theory, one can expect a wide variety of individual
preferences on this point.   For someone with a different preference,
our results about the structure in mode space remain unchanged, and will
still impact any alternative analysis performed to express different
prejudices about possible outcomes of future experiments.  We feel we have
built assumptions into our analysis which allow us to most directly
address the question ``if one of these experiments is carried out and gets a
significant signal, what are the prospects for that signal completely
ruling out a given scalar field model?''.  While we have focused our
quantitative analysis on cosmologies based on one or the other of our
scalar field models, one also can see (by inspecting the mode space
figures) that it would be quite possible to get an experimental signal
that rules out all the scalar field models considered here.  That again
is a consequence of the mode space structure demonstrated in this
paper.

\section{Conclusions}
\label{sec6}

We have considered the ability of future dark energy experiments to
discriminate among different scalar field quintessence models of dark
energy. To this end we have projected the equation of state functions
$w(a)$ for each model into the space of best measured eigenmodes of
future experiments.  We believe this approach is
effective and convenient for investigating the ability of a given data
set to discriminate among different quintessence models. Specifically,
this approach offers a way around the fact that parameters of different quintessence
models are typically not defined in the same spaces, which makes more
direct comparisons of the models problematic.

The four quintessence models considered here create a distinctive
structure when projected into the mode spaces.  The goal of
discriminating among these quintessence models gives an alternative
and complementary measure of the impact of future experiments.  In
large part due to the structure in mode space, this measure has some
striking features that are different from other measures considered
previously.

We have shown that the DETF Stage 3 data will have very little
utility in discriminating among the four quintessence models, although
it will significantly probe the possibility of
non-cosmological-constant-like behavior. DETF Stage 4 simulated data
appears to lie right at an interesting threshold in that this data
shows significant discriminating power among the quintessence models
we considered.  We also showed that modest improvements over DETF
Stage 4 (which many consider quite realistic for some Stage 4
experiments currently proposed) allows one to cross this threshold
more completely, leading to substantially greater discriminating
power.

It is important to note that at our current level of theoretical
understanding all quintessence models are suspect, and we are not
advocating the use of the measures presented here to the
exclusion of other approaches.  However, as discussed in
\cite{Barnard:2007ta,Abrahamse:2007te,Bozek:2007ti,Yashar:2008xx}, we
have chosen an interesting sampling of reasonably motivated
quintessence models. Since such quintessence models are part of
the current theoretical discussion of dark energy, discriminating
power among these models should be part of how we evaluate the
impact of dark energy experiments.

\begin{acknowledgments}

We thank Lloyd Knox for helpful discussions.
We thank Tony Tyson and his group (especially Perry Gee and Hu Zhan) for use of their
computational resources and technical support.
We thank Matthew Auger for his assistance on our figures.
This work was supported in part by DOE grant
DE-FG03-91ER40674 and NSF grant AST-0632901.
\end{acknowledgments}

\bibliography{bib01}
\pagebreak

\end{document}